\documentstyle[12pt]{article}
\begin{document}
\begin{titlepage}

\begin{center}
{\LARGE  Boundary terms in Nambu--Goto string action$^{\star}$}

\vskip 1cm
{\large\bf Leszek Hadasz$^{^{\dag}}$ } \\

{\large\bf Pawe\l~W\c{e}grzyn$^{^{\ddag}}$ } \\
\vskip 1cm

Jagellonian University, Institute of Physics \\
Reymonta 4, 30--059 Krak\'ow, Poland
\end{center}

\begin{abstract}

We investigate classical strings defined by the Nambu-Goto action
with the boundary term added. We demonstrate that the latter term
has a significant bearing on the string dynamics. It is confirmed
that new action terms that depend on higher order derivatives of
string coordinates cannot be considered as continuous perturbations
from the starting string functional. In the case the boundary term
reduces to the Gauss-Bonnet term, a stability analysis is performed
on the rotating rigid string solution. We determine the most generic
solution that the fluctuations grow to. Longitudinal string
excitations are found. The Regge trajectories are nonlinear.

\end{abstract}

\vskip 2cm
\begin{tabular}{l}
TPJU--15/94 \\
August 1994
\end{tabular}

\vspace{\fill}

\noindent
\underline{\hspace*{10cm}}

\noindent
$^{\star}$Work supported in part by grant KBN 2 P302 049 05.

\noindent
$ ^{\dag}$E--mail: hadasz@ztc386a.if.uj.edu.pl

\noindent
$^{\ddag}$E--mail: wegrzyn@ztc386a.if.uj.edu.pl

\noindent
\end{titlepage}

A simple non--interacting string sweeps out  a time--like surface 
of minimal area in four--dimensional Minkowski spacetime. The
minimal surface can be parametrized in such a way that non--linear
equations of motion turn into linear wave equations, and the string
model becomes mathematically tractable. It was found \cite{pawel}
that the most general model of strings, which gives critical
worldsheets of minimal area, is defined by the following action:
\begin{equation}
\label{act}
S = -\gamma A - \frac{\alpha}{2}S_{GB} - \beta S_{Ch},
\end{equation}
where $\gamma$ is string tension, $A$ denotes worldsheet area,
$\alpha$ and $\beta$ are dimensionless parameters 
$(\gamma, \alpha > 0).$  $S_{GB}$ and $S_{Ch}$ are pseudoeuclidean
Gauss--Bonnet and Chern terms (related to Euler characteristics and
surface self--intersections in Euclidean geometry).

String equations of motion derived from (\ref{act}) are 
Nambu--Goto equations supplemented by some edge conditions, which
depend on the action parameters.

The string action (\ref{act}), that depends on two arbitrary
dimensionless parameters, represents a generic form allowed by
reparametrization and Poincare symmetries \cite{pawel} and the
requirement that the variational problem results in minimal surfaces.
This statement is true as long as we do not consider additional 
objects that could couple to strings, like internal fields living
on the worldsheet or constant external fields in the target 
spacetime. Obviously, there exist also ``point--like'' actions,
being functionals of the trajectories of string endpoints. The
simplest example is given by [2--6]:
\begin{equation}
\label{pact}
S_p = -mL_1 -mL_2,
\end{equation}
where $L_1$ and $L_2$ are invariant lengths of the trajectories of
string ends. Such ``non--stringy'' terms modify edge conditions in
the variational problem for critical string worldsheets, but they
cannot be represented as reparametrization invariant surface terms
and do not modify local distributions of energies, momenta and angular
momenta along strings.

The choice of the worldsheet parametrization can be defined by the
following conditions \cite{nest}:
\begin{eqnarray}
\label{param}
(\dot{X} \pm X')^2 & = & 0, \nonumber \\
(\ddot{X} \pm \dot{X}')^2 & = &  -\frac{1}{4}q^2,
\end{eqnarray}
where the dot and the prime stand for derivatives of worldsheet 
coordinates $X_{\mu}(\tau,\sigma)$ with respect to internal string
parameters $\tau$ and $\sigma.$ The parameter $q$ can be considered
as a momentum scale unit. This parameter is freely adjustable.
 In the above parametrization, bulk
equations of motion get linearized and their general solution reads:
\begin{equation}
\label{sol}
X_{\mu}(\tau,\sigma) = X_{L\mu}(\tau+\sigma) + X_{R\mu}(\tau-\sigma).
\end{equation}

To solve boundary problem at string ends $\sigma=\pm\frac{\pi}{2},$
we make use of the correspondence between minimal surfaces $X_\mu$
parametrized according to (\ref{param}) and solutions $\Phi$ of
the complex Liouville equation:
\begin{equation}
\label{liouv}
\ddot{\Phi} - \Phi'' = 2q^{2}e^{\Phi}.
\end{equation}

Any solution of either Nambu--Goto equations together with 
(\ref{param}) or complex Liouville equation (\ref{liouv}) can be
presented in the following form:
\begin{eqnarray}
\label{solution}
e^{\Phi} & = & -\frac{4}{q^2}
\frac{f_{L}'(\tau+\sigma)f_{R}'(\tau-\sigma)}
     {[f_{L}(\tau+\sigma)-f_{R}(\tau-\sigma)]^2} \ , \nonumber \\
\dot{X}^{\mu}_{L,R} & = &
\frac{1}{4|f_{L,R}'|}(1+|f_{L,R}|^2,~2Re\,f_{L,R},~2Im\,f_{L,R},
                      ~1-|f_{L,R}|^2) \ ,
\end{eqnarray}
where $f_L$ and $f_R$ are arbitrary complex functions. 

As the derivatives of left-- and right--movers are light--like
vectors, we can interprete $f_L$ and $f_R$ as their coordinates on
the complex plane, on which the stereographical projection of the
sphere of null directions in four--dimensional spacetime has been
performed. Let us also note that modular transformations of
$f_{L,R}$ induce Lorentz transformations of worldsheet coordinates 
while  Liouville field $\Phi$ remains unchanged.

Now, we can present edge conditions following from the string action
(\ref{act}) as:
\begin{eqnarray}
\label{econd}
e^{\Phi} & = & -\frac{1}{q}\sqrt{\frac{\gamma}{\alpha}}
                e^{-i\theta} \nonumber \ , \\
\label{ec2}
Im \, \Phi' & = & 0 ~~~ \mbox{for}~~ \sigma = \pm\frac{\pi}{2} \ ,
\end{eqnarray}
where the angle parameter $\theta\in[-\pi,\pi]$ is defined by:
\begin{equation}
\label{thdef}
\tan\frac{\theta}{2} = \frac{\beta}{\alpha} \ .
\end{equation}
In this paper we consider only the case $\theta=0$ $(\beta=0).$
This model has been investigated earlier in papers 
\cite{barb1,zhe,barb3}.
Then, there exists a well known solution corresponding to a rotating
rigid rod,
\begin{equation}
\label{rod}
X^{\mu} = \frac{q}{\lambda^2}(\lambda\tau,
~\cos\lambda\tau\sin\lambda\sigma,
~\sin\lambda\tau\sin\lambda\sigma,~0) \ ,
\end{equation}
where the angular frequency $\lambda$ satisfies the relation:
\begin{equation}
\label{lambda}
\frac{\lambda^2}{\cos^2\frac{\lambda\pi}{2}} =
q\sqrt{\frac{\gamma}{\alpha}} \ .
\end{equation}
Note that the velocity of string ends is smaller than the velocity 
of light and tends to it in the limit $\alpha\rightarrow 0$
$(\lambda\rightarrow 1).$

We can compute the energy and the angular momentum of the rotating 
string (\ref{rod}) (for relevant general formulae see ref. 
\cite{pawel}),
\begin{eqnarray}
\label{enmom}
E ~\equiv~ P^0 & = & \frac{\gamma q\pi}{\lambda}\left(
 1 + \frac{\sin\pi\lambda}{\pi\lambda}\right), \nonumber \\
J ~\equiv~ M^{12}  & = & \frac{\gamma q^2\pi}{2\lambda^3}\left[
 1 + 2\frac{\sin\pi\lambda}{\pi\lambda} +
 \frac{\sin2\pi\lambda}{2\pi\lambda}\right]. 
\end{eqnarray}
The total momentum and other components of the total angular momentum
vanish.

The pertinent Regge trajectory is plotted in Fig.1. 
Regge trajectories represent the angular momentum $J$ versus the
squared mass $E^{2}$ relationships for given string configurations.
We have compared trajectories  for a rotating rigid rod
obtained  $(a)$ in the standard Nambu--Goto 
open string model and $(b)$ for the Nambu--Goto string with 
massive ends (due to the point--like terms (\ref{pact}) --- see ref.
\cite{chod}). Asymptotically, in the region  of large masses, the 
trajectory can be approximated by the formula:
\begin{equation}
\label{traj}
J = \frac{1}{2\pi\gamma}E^2 + \frac54\left(\frac{\alpha}
{\pi^6\gamma^3}\right)^{1/4}E^{3/2}\ .
\end{equation}

We see that it is slightly raised  in comparison with the 
Nambu--Goto open string trajectory. At low masses, unlike the case
$(b)$ where the appearance of point like masses at string ends 
curves the trajectory downwards and the intercept is lowered, we
find here approximately a linear dependence:
\begin{equation}
\label{low}
J = \sqrt{\frac{\alpha}{\gamma}}E\ .
\end{equation}
It is interesting to note that the energy distribution along the
string has been also changed. For the Nambu--Goto rotating rigid 
string  the energy density is constant. In the modified model,
the energy density (plotted in Fig.2) is given by the formula:
\begin{equation}
\label{enden}
p^0 = \frac{\gamma q}{\lambda}\left[1+\cos^4\frac{\lambda\pi}{2}
\left(\frac{3}{\cos^4\lambda\sigma}-\frac{2}{\cos^2\lambda\sigma}
\right)\right].
\end{equation}
We now turn to study small fluctuations around the solution
(\ref{rod}). This solution is associated to a static solution
of Liouville equation (\ref{liouv}),
\begin{equation}
\label{zero}
e^{\Phi_0} = - \frac{1}{q^2}\frac{\lambda^2}{\cos^2\lambda\sigma}.
\end{equation}
A small perturbation $\Phi_1$ from the static solution $\Phi_0$ 
satisfies the following linear equation:
\begin{equation}
\label{leq}
\ddot{\Phi}_1 - \Phi_1'' = -V(\sigma)\Phi_1\ ,
\end{equation}
where we have denoted
\begin{equation}
\label{pot}
V(\sigma) = \frac{2\lambda^2}{\cos^2\lambda\sigma}\ .
\end{equation}
The solution $\Phi_1$ is subject to the following boundary conditions
at $\sigma = \pm\frac{\pi}{2}:$
\begin{eqnarray}
\label{bc}
\Phi_1 & = & 0\ , \\
\label{bc2}
Im \, \Phi_1' & = & 0\ .
\end{eqnarray}
One can prove that the imaginary part of $\Phi_1$ must vanish
at any worldsheet point. Thus, the Liouville field  $\Phi_1$ is real.
We can separate variables to find a solution satisfying 
the equation (\ref{leq}) together with the boundary conditions
(\ref{bc}):
\[
\Phi_1(\tau,\sigma) = T(\tau)\Sigma(\sigma)\ .
\]
We obtain a system of ordinary differential equations:
\begin{eqnarray}
\label{e18}
\ddot{T} + {\cal E}T & = & 0\ , \\
\label{eb18}
\left(-\frac{d^2}{d\sigma^2} +V(\sigma)\right)\Sigma & = & 
{\cal E}\Sigma\ ,
\end{eqnarray}
together with boundary conditions:
\begin{equation}
\label{bc18}
\Sigma = 0 ~~~ \mbox{for} ~~~ \sigma = \pm\frac{\pi}{2}.
\end{equation}
The solutions of Schr\"odinger equation (\ref{eb18}) that obey
periodic boundary conditions (\ref{bc18}) can exist only if 
${\cal E} > 2\lambda^2,$  where $2\lambda^2$ is the minimal value
of the potential $V(\sigma).$ It implies that the separation
constant $\cal E$ must be positive. For convenience we introduce new
variable $\omega$ defined as:
\[
{\cal E } = \omega^2\ .
\]
The Schr\"odinger equation (\ref{eb18}) with the potential 
(\ref{pot}) can be exactly solved. The solutions exist only for 
discrete  values of the separation constant $\omega = \omega_n \ 
(n=1,2,\ldots)$, being roots of the following equation (see Fig.3):
\begin{equation}
\label{omegi}
\omega_n\tan\frac{\pi(\omega_n+n)}{2} =
\lambda\tan\frac{\pi\lambda}{2}\ .
\end{equation}
The general solution of the equation (\ref{leq}) satisfying boundary
conditions (\ref{bc}, \ref{bc2}) reads:
\begin{equation}
\label{eq20}
\Phi_1 = 2\sum_{n/1}^{\infty}D_n\sin(\omega_n\tau+\varphi_n)
 \left[\tan\lambda\sigma\cos\left(\omega_n\sigma+
 \frac{n\pi}{2}\right) -                     \\
 \frac{\omega_n}{\lambda}\sin\left(\omega_n\sigma+
 \frac{n\pi}{2}\right)\right],
\end{equation}
where $D_n$ and $\varphi_n$ are arbitrary real constants.

To visualize string worldsheets that correspond to Liouville fields
$\Phi = \Phi_0 + \Phi_1$ we must employ the relations (\ref{solution}).
Taking into account that $e^\Phi$ is real, we can make functions 
$f_L$ and $f_R$ unimodular (by some modular transformation --- it
is equivalent to specifying some reference frame in Minkowski
spacetime). Then, it is convenient to introduce new real fields 
$F_L$ and $F_R,$ 
\[
f_{L,R} = e^{iF_{L,R}}\ ,
\]
and the relations (\ref{solution}) go over into:
\begin{eqnarray}
\label{ea21}
e^{\Phi} & = & -\frac{1}{q^2}\frac{F_L'F_R'}{\sin^2\frac{F_L -
                F_R}{2}}\ , \\
\label{eb21}
\dot{X}^{\mu}_{L,R} & = & \frac{1}{2F_{L,R}'}(1,~\cos F_{L,R},
                          ~\sin F_{L,R},~0)\ .
\end{eqnarray}
The static field $\Phi_0$ corresponds to:
\begin{equation}
\label{eq22}
F^{(0)}_L = \lambda(\tau + \sigma), ~~~
F^{(0)}_R = \lambda(\tau - \sigma) + \pi,
\end{equation}
while the first order fluctuations $\Phi_1$ are associated to the 
following corrections:
\begin{equation}
\label{eq23}
F^{(1)}_{L,R} = \pm\sum_{n/1}^{\infty}D_n\sin\left[\omega_n(\tau
\pm \sigma) + \varphi_n \pm \frac{n\pi}{2}\right],
\end{equation}
where plus and minus signs correspond to left-- and right--movers
respectively.

In contrast to the Nambu--Goto case, there appear longitudinal
excitations of the string. Moreover, only such kind of fluctuations
come out at the first order. With the help of formulae above, the total
string length $L$ can be evaluated at some fixed time $X^0:$
\begin{eqnarray}
\label{lenght}
\frac{\lambda^2L}{2q} & = & \sin\frac{\pi\lambda}{2} -
        \sum_{n/1}^{\infty}\frac{D_n}{\omega_n^2 - \lambda^2}
        \sin\left(\frac{\lambda\omega_n}{q}X^0 + \varphi_n
        \right)\times   \nonumber \\
  &   & \left[\omega_n^2 + \lambda^2 + 2\lambda^2\tan^2\left(
        \frac{\pi\lambda}{2}\right)\right]
        \cos\left[\frac{\pi(\omega_n + n)}{2}\right]
        \cos\left(\frac{\pi\lambda}{2}\right).
\end{eqnarray}

Let us now calculate the contribution to the energy coming from
fluctuations. The general formula for the total string energy reads:
\begin{equation}
\label{eq24}
P^0 = \frac{\gamma q}{2}\int_{-\frac{\pi}{2}}^{\frac{\pi}{2}}\!\!
      d\sigma\left(\frac{1}{F_L'}+\frac{1}{F_R'}\right) -
      \frac{\gamma q}{2}\left[\frac{\sin(F_L - F_R)}{F_L'F_R'}
      \right]^{\sigma=+\frac{\pi}{2}}_{\sigma=-\frac{\pi}{2}}.
\end{equation}
The straightforward calculations lead to the following result:
\begin{eqnarray}
\label{eq25}
P^0 & = & \frac{\gamma q\pi}{\lambda}\left(1+
          \frac{\sin\pi\lambda}{\pi\lambda}\right) \nonumber \\
 & & +~ \frac{\gamma q\pi}{2\lambda^3}\sum_{n/1}^{\infty}
     D_n^2\omega_n^2\left[1+\frac{\sin\pi\lambda}{\pi\lambda}+
     2(-1)^{n+1}\cos\pi\lambda\frac{\sin\pi\omega_n}{\pi\omega_n}
     \right].
\end{eqnarray}
One can easy check that the energy of fluctuations  is always
positive. It means that the solution (\ref{rod}) is stable against
small perturbations.
In fact, to calculate the total string energy (\ref{eq25}) up to the
second order we need also to consider second order corrections to
the zero order solution. It can be proved by straightforward
calculations that they do not contribute to the energy at the second
order.

Finally, we want to summarize our results.
We examined a classical string model in four--dimensional 
Min\-kow\-ski
space\-time defined by the Nambu--Goto action with some boundary term
added. It warrants that critical worldsheets are minimal surfaces, 
but some non--linear, third order in time derivatives equations hold
at string ends. It is evident from this paper that additional terms
to the action functional depending on the second order derivatives
of string coordinates cannot be regarded as higher order corrections
to the starting Nambu--Goto action. In the limit of vanishing
coupling constants $(\alpha,\beta \to 0)$ our model does not 
revert to the original Nambu--Goto string model. There are still
higher order equations (\ref{ec2}) to be satisfied. This is an 
unavoidable consequence of employing theoretical framework for string
actions with second order derivatives. The number of boundary
conditions for dynamical equations of motion is doubled. The same is 
true for the number of initial data necessary to formulate properly 
the Cauchy problem. Roughly speaking, passing to dynamical models, 
that are governed by the variational principle with  actions 
depending on second order derivatives of dynamical variables, doubles
the number of independent degrees of freedom. 

A generic minimal wordsheet model (\ref{act}) has been investigated
for $\beta = 0$. We have found a classical ground state solution, 
that corresponds to a rotating rigid rod. Unlike for the analogous
Nambu--Goto configuration, string ends move with the velocity smaller
than the velocity of light and non--relativistic limit can be 
defined. It has been shown that the mass distribution along the 
string has been changed. Regge trajectories in this model are
non--linear. The ground state solution is stable against small 
perturbations. Eigenfrequencies for each fluctuation mode are found
to be solutions of some simple transcendental equation.
The excitations give a positive contribution to the
total energy of the string. Another interesting property is that
perturbations do not disturb the string from the planar motion,
the shape of the string lies in a plane. But its total length 
measured in the laboratory frame oscillates, in contrast to other
 classical string models \cite{nest2}.

This work was supported in part by the KBN under grant
2 P302 049 05.

\begin{flushleft}
{\large {\bf Figure Captions}} \\

\vskip .5cm

\begin{tabular}{ll}
Fig. 1. &
Regge trajectories for various string models: \\
  &
(a) Nambu--Goto string, \\
  &
(b) Rebbi--Thorn--Chodos string with massive ends, \\
  &
(c) string model with Gauss--Bonnet boundary term. \\
 & \\
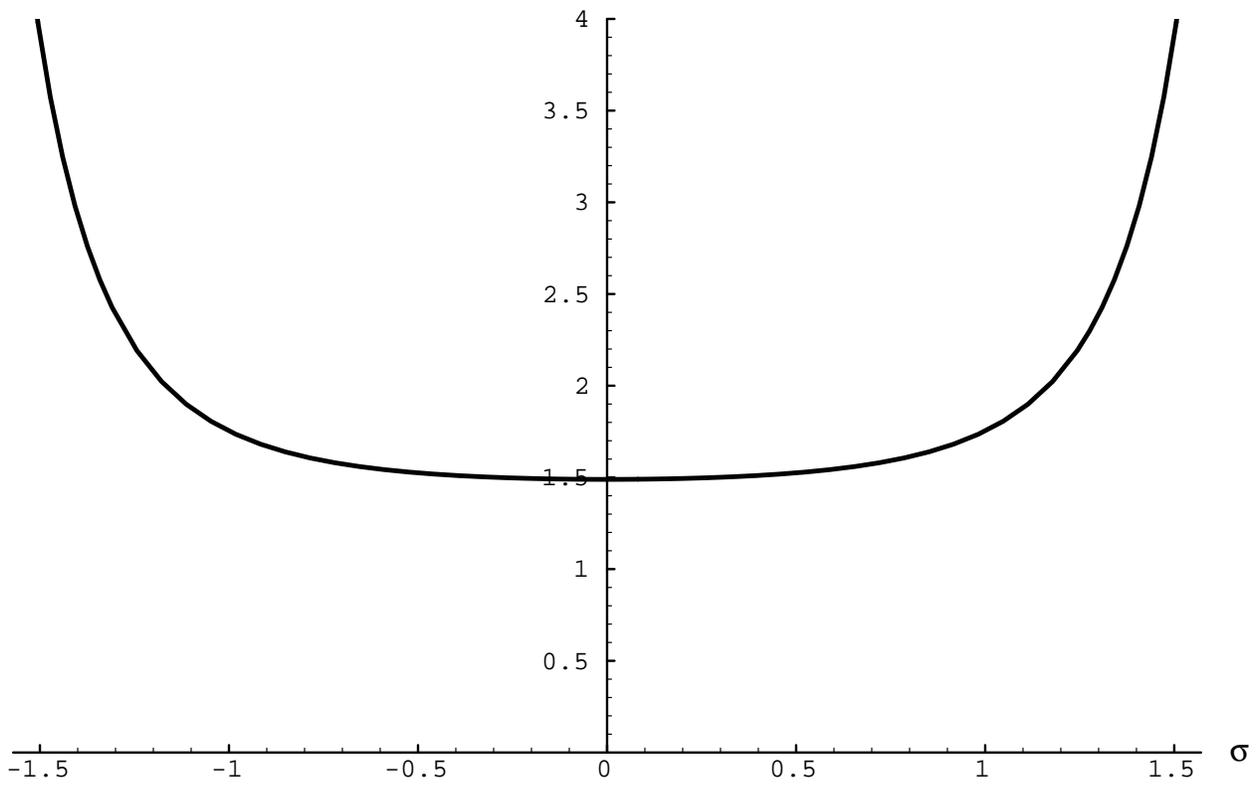
Fig. 2. &
The shape of energy density distribution along the string. \\
 & \\
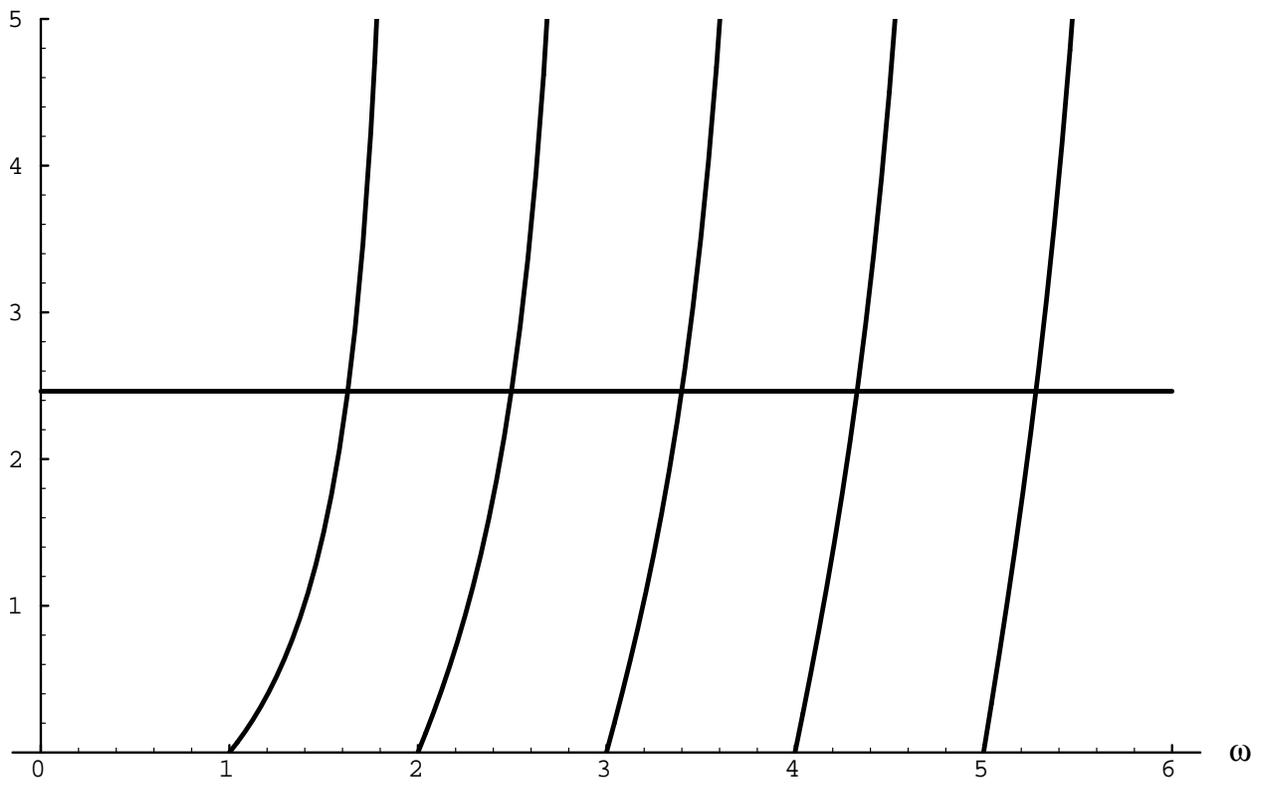
Fig. 3. &
Graphical solution of the equation 
$\lambda\tan\left(\frac{\pi\lambda}{2}\right) =
\omega\tan\left[\frac{\pi(\omega+n)}{2}\right]$ \\
 & for eigenfrequencies $\omega$ of fluctuation modes. 
The parameter $\lambda$ \\
 & is a fixed value of angular frequency
the rigid rod rotates with.
\end{tabular}
\end{flushleft}

\end{document}